\documentclass[final]{aipproc}

\layoutstyle{6x9}

\newbox\grsign \setbox\grsign=\hbox{$>$} \newdimen\grdimen
\grdimen=\ht\grsign
\newbox\simlessbox \newbox\simgreatbox \newbox\simpropbox
\setbox\simgreatbox=\hbox{\raise.5ex\hbox{$>$}\llap
     {\lower.5ex\hbox{$\sim$}}}\ht1=\grdimen\dp1=0pt
\setbox\simlessbox=\hbox{\raise.5ex\hbox{$<$}\llap
     {\lower.5ex\hbox{$\sim$}}}\ht2=\grdimen\dp2=0pt
\setbox\simpropbox=\hbox{\raise.5ex\hbox{$\propto$}\llap
     {\lower.5ex\hbox{$\sim$}}}\ht2=\grdimen\dp2=0pt
\def\simgreat{\mathrel{\copy\simgreatbox}}
\def\simless{\mathrel{\copy\simlessbox}}

\begin{document}

\title{New white dwarfs for the stellar initial mass-final mass relation}


\author{Paul D Dobbie}{
  address={Australian Astronomical Observatory, Epping, NSW 1710}
}

\author{Richard Baxter}{
  address={Dept. of Physics \& Astronomy, Macquarie University, NSW 2109}
}

\begin{abstract}

We present the preliminary results of a survey of the open clusters NGC3532 and NGC2287 for new white
dwarf members which can help improve understanding of the form of the upper end of the stellar initial
mass-final mass relation. We identify four objects with cooling times, distances and proper motions 
consistent with membership of these clusters. We find that despite a range in age of $\sim$100Myrs the
masses of the four heaviest white dwarfs in NGC3532 span the narrow mass interval M$_{WD}$$\approx$0.9$-$1.0M$_{\odot}$ suggesting that
the initial mass-final mass relation is relatively flatter over 4.5M$_{\odot}$ $\simless$M$_{init}$
$\simless$6.5M$_{\odot}$ than at immediately lower masses. Additionally, we have unearthed WD J0646-203
which is possibly the most massive cluster white dwarf identified to date. With M$_{WD}$$\approx$1.1M$_{\odot}$ 
it seems likely to be composed of ONe and has a cooling time consistent with it having evolved from a single 
star. 
 
\end{abstract}

\maketitle


\section{Probing the fate of heavy-weight intermediate mass stars}

Stars with masses, M$\simless$5M$_{\odot}$ (ie. $>$95\% of the stellar population by number) will ultimately become white dwarfs, with electron degenerate cores 
composed of He or C and O. Those much rarer stars with M$\simgreat$10M$_{\odot}$ will burn their non-degenerate cores through to Fe and then die as core-collapse Type 
II supernovae. However, the final evolutionary fate of single stars in the intervening mass range remains less certain. Early stellar evolutionary models predicted 
that when the partially degenerate CO core of a heavy-weight intermediate mass early-asymptotic giant branch (E-AGB) star achieved M$\sim$1.1M$_{\odot}$ it ignited and burned 
to Ne and O before collapsing. The star was then expected to explode as an electron capture Type II supernova (e.g. \cite{nomoto84}). However, more modern calculations which include improved physics suggest that a sizeable proportion of the stars in this initial mass range may instead pass through a super-AGB phase before ending 
their lives as ultra-massive, M$\sim$1.05-1.3M$_{\odot}$, ONe white dwarfs (e.g. \cite{garcia97}). 

Refining our knowledge of the fate of stars in this initial mass range is important for understanding the chemo-dynamical evolution of the Galaxy. Supernovae inject substantial amounts of kinematic energy into the ISM (e.g. \cite{cox74}). Moreover, due to the highly compact nature of the core during the super-AGB phase, the convective envelopes of these stars are subject to intense hot bottom burning and consequently they synthesise substantial amounts of $^{14}$N and $^{13}$C which are returned to the ISM (e.g. \cite{siess10}). Here we report preliminary results from our search of the moderately rich and relatively nearby open clusters NGC2287 and NGC3532 for the white dwarf remnants of stars which are believed to have had initial masses within this range of interest.

\begin{figure}
  \includegraphics[height=.6\textheight,angle=270]{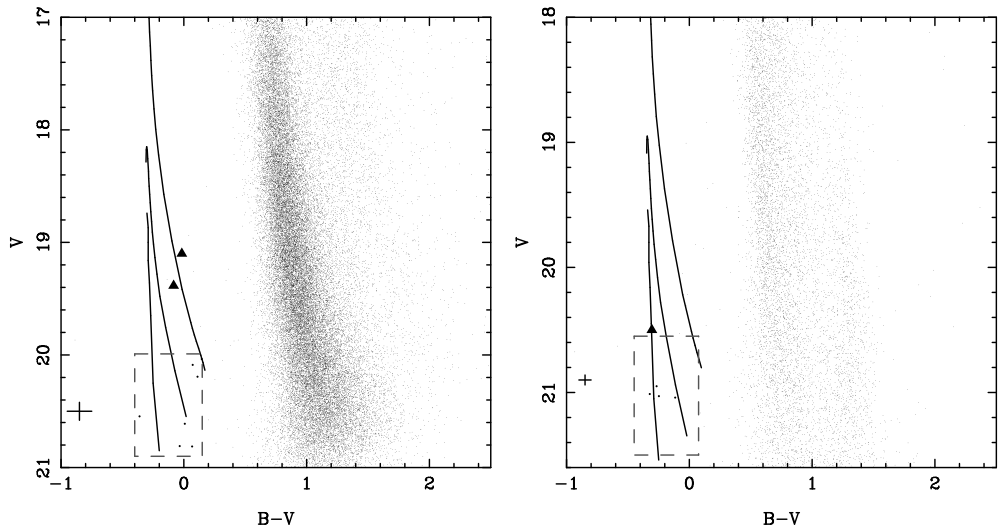}
  \caption{$V$, $B$-$V$ colour-magnitude diagrams for NGC3532 (left) and NGC2287 (right). Theoretical evolutionary tracks for 0.7M$_{\odot}$ and 1.1M$_{\odot}$ 
CO core \cite{holberg06} and 1.16 ONe core \cite{althaus07} white dwarfs, adjusted to account for the 
foreground reddening towards and the distance of each population, are overplotted. Previously known white dwarf members of 
the clusters recovered here are also highlighted (large triangles). }
\end{figure}

\section{Newly identified white dwarf cluster members}

\subsection{Survey imaging}

We retrieved $B$ and $V$ band imaging of the open clusters NGC3532 and NGC2287 from the ESO archive. These observations 
were obtained with the 2.2m telescope and the Wide Field Imager on the nights of 1999/12/04 and 2000/02/24 respectively.
The images were reduced following standard methods applied using the Cambridge Astronomical Survey Unit CCD reduction 
toolkit \cite{irwin01}. We performed aperture photometry 
on the reduced frames and used the limited standard star observations from these nights to crudely calibrate these data.

A $V$, $B$-$V$ colour-magnitude diagram was constructed for each open cluster (Figure 1). As a guide to the likely location of the 
white dwarf sequence in NGC2287 and NGC3532, we overplotted theoretical evolutionary tracks for 0.7M$_{\odot}$ and 1.1M$_{\odot}$ 
CO core \cite{holberg06} and 1.16 ONe core \cite{althaus07} degenerates, which were adjusted to account for the 
foreground reddening towards and the distance of each population.  We identified all objects classified as stellar lying within 
the boxes overplotted in Figure 1 as candidate white dwarf cluster members.


\begin{table}
\begin{tabular}{cccc}
\hline
 ID &$T$$_{\rm eff}$(K) & log $g$ &  m$_{V}$     \\
\hline

 WD J1105-585 & $13329^{+647}_{-442}$ & $8.19^{+0.06}_{-0.11}$ & 20.29$\pm$0.04  \\ 
 WD J1106-590 & $19690^{+333}_{-339}$ & $8.45^{+0.03}_{-0.03}$ & 20.01$\pm$0.04 \\ 
 WD J1106-584 & $18833^{+338}_{-335}$ & $8.52^{+0.03}_{-0.03}$ & 20.12$\pm$0.04 \\ 
 WD J1107-584 & $20923^{+505}_{-497}$ & $8.61^{+0.07}_{-0.07}$ & 20.16$\pm$0.04 \\

 WD J0644-205 & $12376^{+230}_{-233}$ & $7.98^{+0.06}_{-0.05}$ & 20.52$\pm$0.04 \\ 
 WD J0645-202 & $14716^{+378}_{-452}$ & $7.97^{+0.04}_{-0.04}$ & 20.98$\pm$0.04 \\ 
 WD J0646-203 & $25520^{+380}_{-351}$ & $8.82^{+0.06}_{-0.06}$ & 20.91$\pm$0.04 \\ 

\hline
\end{tabular}
\caption{Observed properties of the seven new white dwarf candidate members of NGC3532 and NGC2287 observed with the VLT.}
\end{table}

\subsection{Spectroscopic follow-up}

We used the Very Large Telescope (VLT) and FORS instrument on the nights of 2010/02/06-07 to spectroscopically observe three and four
candidates in NGC2287 and NGC3532 respectively. These datasets were reduced following standard procedures within the IRAF 
software environment. Observations of a He+HgCd arc lamp were used to wavelength calibrate the extracted datasets and observations 
of the DC white dwarf LHS2333 were used to remove residual features in the data arising from the spectral response of the system. 

All seven candidates have spectra consistent with DA white dwarfs. To measure the effective temperatures and surface gravities of 
these objects, as in our previous work \cite{dobbie09a,dobbie09b} we compared the H-$\beta$ $-$ H-8 Balmer lines in the 
observed spectrum of each star to a grid of synthetic line profiles generated with the model atmosphere codes TLUSTY and SYNSPEC. 
The results of this analysis are shown in Table 1. Subsequently, the masses and the cooling times of the white dwarfs have been 
determined using the CO core, thick-H layer evolutionary models of \cite{fontaine01}. These are displayed in Table 2. 

We find the cooling times of WD J0644-205 and WD J1105-585 are greater than the age of their proposed host populations (NGC2287,
$\tau$$\sim$250Myrs; NGC3532, $\tau$$\sim$300Myrs) and thus conclude that these are unlikely to be associated with the clusters.

\begin{table}
\begin{tabular}{ccccc}
\hline
ID & M$_{V}$ & M(M$_{\odot}$)  &  $\tau_{c}$ (Myrs) & M$_{\rm init}$(M$_{\odot}$) \\
\hline

 WD J1105-585 & $11.77\pm{+0.11}$ & $0.72\pm0.04$ & $370^{+42}_{-37}$  &  - \\ 
 WD J1106-590 & $11.51\pm{+0.13}$ & $0.90\pm0.04$ & $188^{+26}_{-23}$   &  5.15$^{+0.61}_{-0.37}$\\ 
 WD J1106-584 & $11.71\pm{+0.13}$ & $0.94\pm0.04$ & $243^{+32}_{-28}$    &  6.93$^{+3.70}_{-1.21}$\\ 
 WD J1107-584 & $11.69\pm{+0.13}$ & $1.00\pm0.04$ & $210^{+29}_{-26}$    &  5.64$^{+1.18}_{-0.56}$\\ 

 WD J0644-205 & $11.59\pm{+0.10}$ & $0.59\pm0.04$ & $334^{+34}_{-30}$   &  - \\ 
 WD J0645-202 & $11.27\pm{+0.11}$ & $0.59\pm0.04$ & $200^{+23}_{-21}$    &  - \\ 
 WD J0646-203 & $11.73\pm{+0.15}$ & $1.12\pm0.04$ & $173^{+25}_{-21}$    &  6.32$^{+1.34}_{-0.76}$\\

\hline
\end{tabular}
\caption{Derived properties of the seven new white dwarf candidate members of NGC3532 and NGC2287 observed with the VLT.
Errors in absolute magnitudes, masses and cooling times shown here are derived by propagating more realistic
uncertainties in effective temperature and surface gravity of 2.3\% and 0.07dex respectively.}
\end{table}

\subsection{Distances and proper motions}

We obtained new $V_{\rm }$  band imaging of the five remaining white dwarf candidate cluster members with the Inamori Magellan Areal Camera
and Spectrograph (IMACS) and the Magellan Baade telescope on the photometric night of 2010/04/09. To photometrically
calibrate these images and transform our $V$ band instrumental magnitudes onto the Johnson system, we also observed a number of different 
standard star fields throughout the course of the night. The frames were reduced using the Cambridge Astronomical Survey Unit CCD reduction 
toolkit (\cite{irwin01}) to follow standard procedures. As before, aperture photometry was performed on the reduced images using a circular
window with a diameter of 1.5$\times$ the full width half maximum of the mean point spread function. The measured $V$ magnitudes for the white 
dwarfs are shown in Table 1.

We have used our spectroscopic effective temperature and surface gravity measurements and grids of synthetic photometry for DA white dwarfs 
\cite{holberg06} to determine absolute $V$ band magnitudes (Table 2). Subsequently, we used these in conjunction with the observed $V$ magnitudes to 
estimate the distances of the remaining white dwarf candidate cluster members. We determine that WD J1106-590, WD J1106-584 and WD J1107-584 
have distances consistent with that of NGC3532. Additionally, our distance estimate for WD J0646-203 is consistent with that of NGC2287. However,
we find that WD J0645-202 lies $\simgreat$150pc behind NGC2287 and thus conclude that it is unlikely to be associated with this cluster (Figure 2).
 We have also taken advantage of the ten year baseline between the original WFI survey imaging and new IMACS data to measure proper motions. We 
find that our estimates of the tangential velocities of the remaining white dwarf candidate cluster members are consistent with those of their 
proposed host populations.

\section{The new cluster white dwarfs and the stellar IFMR}

In prior work we proposed GD50 as the most massive open cluster white dwarf (M$_{WD}$=1.25M$_{\odot}$) currently known (\cite{dobbie06a}). However, nagging doubts remain about the association of this object with the Pleiades since it resides $\sim$100pc beyond the cluster tidal radius. In contrast, not only does WD J0643-203 have a distance and proper motion consistent with NGC2287, it also lies comfortably within the projected tidal radius of this cluster. With a spectroscopic mass of M$_{WD}$=1.12$\pm$0.04M$_{\odot}$ based on CO core models or  M$_{WD}$=1.08 $\pm$0.04M$_{\odot}$ assuming a ONe core, it is slightly more massive than the Pleiades white dwarf LB1497, M$_{WD}$=1.02$\pm$0.02M$_{\odot}$  (\cite{dobbie06b}) and the heaviest white dwarfs recently identified in NGC2068, M$_{WD}$=1.01-1.02M$_{\odot}$ (\cite{williams09}). Thus WD J0643-203 is potentially the most massive open cluster degenerate identified to date. With a mass that is larger than the theoretical C burning limit of M=1.06M$_{\odot}$ (\cite{nomoto84}), it is a strong contender for an ONe degenerate. Moreover, the cooling age of this object ($\tau_{cool}$$\sim$170Myr) is consistent with it having descended from a single heavy-weight intermediate mass star, lending support to the predictions of modern stellar evolutionary models.

\begin{figure}
  \includegraphics[height=.6\textheight,angle=270]{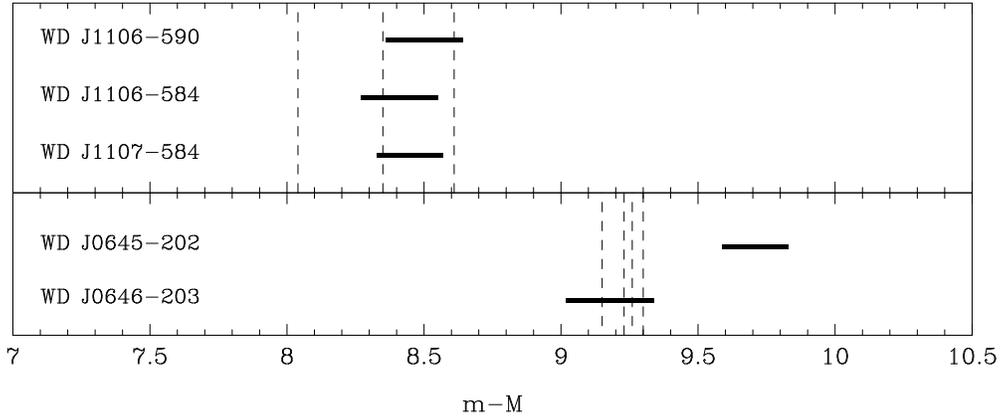}
  \caption{Distance modulii of the white dwarf candidate members of NGC3532 (top) and NGC2287 (bottom). Various distance estimates to each cluster 
are overplotted (dashed vertical lines). }
\end{figure}

Our work has also increased the total number of known white dwarf members of NGC3532 to seven. Since this cluster has an age of $\tau$$\sim$300Myr 
these remnants must be descended from stars with M$_{init}$$\simgreat$3.6M$_{\odot}$. These data are likely to span the initial mass range where the 
stellar initial mass-final mass relation is expected to change gradient due to the onset of the second dredge-up process (e.g. \cite{iben}) and 
provide a better test than white dwarfs drawn from several populations which are affected by cluster to cluster age uncertainties. The locations of 
the seven NGC3532 white dwarfs in initial mass-final mass space, for an assumed cluster age of $\tau$=300Myr, are shown in Figure 3 (also see Table 2). After matching both a simple line and more complex function allowing for a change in gradient at M$_{init}$$\sim$4M$_{\odot}$ to these data we are able to perform an F-test on the two fit statistics which reveals that the improvement in fit offered by the latter model is modestly significant, P$\sim$0.1. While these data do not provide a decisive demonstration of a change in the gradient of the IFMR at M$_{init}$$\sim$4M$_{\odot}$ they at least offer some further support to our earlier conclusion (\cite{dobbie09b}). A more spatially extensive search of NGC3532 may bolster this result.



\section{Conclusions and future work}

\begin{figure}
  \includegraphics[height=.6\textheight,angle=270]{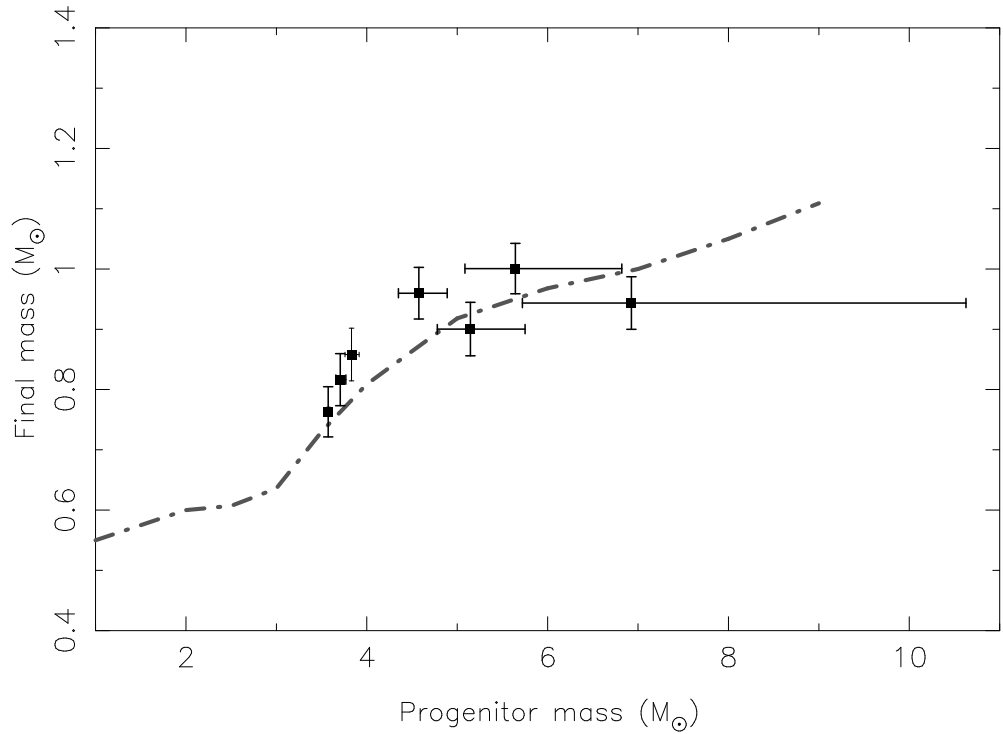}
  \caption{Location of the seven white dwarf members of NGC3532 in initial mass-final mass space for an assumed cluster age of 300Myr. A 
theoretical IFMR \cite{kovetz09} is overplotted (thick grey line).}
\end{figure}

$B$ and $V$ band surveys of NGC2287 and NGC3532 have unearthed four new white dwarfs which are probable members of these two open 
clusters. Somewhat surprisingly, despite a range in effective temperature of $\sim$20\%, corresponding to a range in age of $\sim$100Myr, 
the masses of the four heaviest white dwarf members of NGC3532 all lie between M$_{WD}$$\approx$0.9-1.0M$_{\odot}$. This suggests that there
is little change in remnant mass for initial masses, 4.5M$_{\odot}$ $\simless$M$_{init}$$\simless$6.5M$_{\odot}$, ie. the IFMR is relatively 
flatter here than at immediately lower masses as expected from theory. WD J0646-203 is possibly the most massive cluster white dwarf identified
 to date and with M$_{WD}$=1.12/1.08$\pm$0.04M$_{\odot}$, based on CO/ONe models, seems likely to be composed of ONe. It has a cooling time 
consistent with it having evolved from a single star.   

We have in hand deep imaging for a number of other open clusters which we aim to use to further hone understanding of the form of the IFMR.
NGC752, NGC6940 and NGC2477 will allow us to probe the relatively unexplored region 2M$_{\odot}$M$_{init}$$\simless$3M$_{\odot}$ which includes 
the dividing line between stars that do and do not experience the helium flash. Excitingly, we have $\sim$80 candidate members of NGC2477 
with V$\simless$24.2 which could provide the chance to probe the relative form of the IFMR from M$_{init}$=2M$_{\odot}$ upto the electron capture 
SNe limiting mass in unprecedented detail. Data for the young and extremely rich NGC6791 will allow us to further scrutinise the upper limit on 
the mass of a white dwarf formed via single star evolution.

\end{document}